\documentstyle[amsfonts,aps]{revtex}

\begin{document}
\title{A unified (classical-quantum-statistical) formalism for continuous spectrum
systems.}
\author{Mario Castagnino}
\address{Instituto de Astronom\'{i}a y F\'{i}sica del Espacio.\\
Casilla de Correos 67, Sucursal 28\\
1428, Buenos Aires, Argentina.}
\maketitle

\begin{abstract}
A unified (classical-quantum-statistical) formalism for a system with
continuous spectrum is introduced. For this kind of systems ergodicity
behavior and the existence of microcanonical and canonical (KMS) equilibrium
is proved. It is argued that the continuous spectrum condition is essential
for the thermodynamical behavior.
\end{abstract}

\section{Introduction.}

Systems with continuous spectrum have been studied at the classical,
quantum, and statistical level. In papers \cite{Ant}, \cite{CyLII}, and \cite
{Fund} we have contributed to this research. Precisely, based on the ideas
of van Hove \cite{vH} we have introduced a method to deal with these
systems, that we have successfully used to treat decaying phenomena,
statistical mechanics problems and to find the classical limit of quantum
systems \cite{Deco}. In this paper, using the same technique, we would like
to present a panoramic version of the different ''limits''\footnote{%
Usually the relations between physical levels are called ''limits'', even
though only in some cases they are just simple mathematical limits.} between
physical levels. We will see that we can use our method (complemented by
other ideas) as a unified formalism that allows to go from one level to the
other. Even if some features of the problem are well known from a long time
we believe that a unified treatment of these matters, like ours, was
missing. Furthermore, most probably, it will be quite useful. Essentially we
believed it is useful to see how the interplay of classical and quantum
concepts can be used to solve problems.

So let us begin looking at the whole panorama. Physics studies two kind of
systems:

A.-{\it \ Individual systems, }like particles or single physical systems.
This is the case in classical mechanics, electromagnetism, special, and
general relativity.

B.-{\it Statistical systems}, like ensembles or sets with many copies of a
single physical system. This is the case in quantum mechanics, statistical
(classical or quantum) mechanics, thermodynamics, and quantum field theory.

In category A the limits between different levels are simple and well known:

A$_1.-$ We can go from general relativity to special relativity if we
consider just flat space-times.

A$_2.-$ In the ''classical limit'' ($\beta =\frac vc\rightarrow 0$) special
relativity becomes classical mechanics.

The limit between the two categories, e. g.: of statistical mechanics
towards classical mechanics is produced by the phenomenon of localization
that we have study in paper \cite{Deco}, where the necessary conditions
under which this phenomenon take place (localizing potential, adequate
initial conditions, etc.) are given for some examples\footnote{%
The study of the localization phenomena is a deep and complicated subject,
since in its classical version is related with the problem of the ''limit
circle'' (see \cite{Splach}, cap II). This is one of the problems that
Hilbert listed for the XX century mathematicians and it remains still
unsolved.}. So we will just make a short comment of this problem in the
paper.

In category B the limits are much more involved and will be considered in
this paper. So it is organized as follows:

B$_{1}.-$ The thermalization of a quantum system is considered in section
II. We will study it in two steps:

B$_{1.1}.-$A first step is decoherence (section II.A). This section, as
section III, it is a brief resume of ref. \cite{Deco} included in the paper
for the sake of completeness. The conditions under which the evolution of
the quantum system decohere (more than one bound state, small decoherence
time, etc.) are listed in the just quoted paper

B$_{1.2}.-$We will see how we can obtain a KMS equilibrium after decoherence
(section II.B).

B$_2.-$ The quantum to classical statistical mechanic limit is studied in
section III. .

B$_{3.-}$ The statistical classical mechanics relation with the theory of
dynamical system and with thermodynamics will be considered in two steps in
section IV:

B$_{3.1}.-$ In the first step the classical statistical system is considered
as dynamical one. Then the micro-canonical ensembles appear as a consequence
of dynamical systems theorems (section\ IV.A).

B$_{3.2.-}$ Finally, in the second step, the canonical ensembles and
thermodynamic are introduced (section IV.B).

Other limit can be obtained combining those above. Knowing all these matters
we can foresee the final fate of a quantum system (to remain a quantum one,
or a classical one, or to end in thermodynamical equilibrium, etc.).

Section V is devoted to the localization problem (namely the $B\rightarrow A$
limit).

We will state our conclusion in section VI.

\section{Quantum to statistical limit.}

We will demonstrate that for a wide set of quantum systems non-diagonal
density operators can be considered as the transient phase while diagonal
density operators is the final regime and a permanent state\footnote{%
The arguments of this section are fully developed in paper \cite{Deco},
where the philosophy of the method is explained at large.}. We will find a
basis where exact density operators decoherence appears for these final
states . Then we will see how the quantum systems thermalize.

\subsection{Decoherence.}

\subsubsection{Decoherence in the energy.}

Let us consider a closed and isolated quantum system with $N+1$ dynamical
variables and a Hamiltonian endowed with a continuous spectrum and just one
bounded ground state. So the discrete part of the spectrum of $H$ has only
one value $\omega _{0}<0$ \footnote{%
The case of many bound states is considered in paper \cite{Deco}. Here we
consider the case with only one bound state because in this case the system
usually decoheres.} and the continuous spectrum is, let say, $0\leq \omega
<\infty .$ Let us assume that it is possible to diagonalize the Hamiltonian $%
H$, together with $N$ observables $O_{i}$ ($i=1,...,N)$. The operators ($H$, 
$O_{1}$,...,$O_{N}$) form a {\it complete set of commuting observables}
(CSCO). For simplicity we also assume a discrete spectrum for the $N$
observables $O_{i}$. Therefore we write 
\begin{equation}
H=\omega _{0}\sum_{m}|\omega _{0},m\rangle \langle \omega
_{0},m|+\int_{0}^{\infty }\omega \sum_{m}|\omega ,m\rangle \langle \omega
,m|d\omega  \label{2.4}
\end{equation}
where $\omega _{0}<0$ is the energy of the ground state, and $m\doteq
\{m_{1},...,m_{N}\}$ labels a set of discrete indexes which are the
eigenvalues of the observables $O_{1}$,...,$O_{N}$. $\{|\omega _{0},m\rangle
,|\omega ,m\rangle \}$ is a basis of simultaneous generalized eigenvectors
of the CSCO: 
\begin{eqnarray*}
H|\omega _{0},m\rangle &=&\omega _{0}|\omega _{0},m\rangle ,\quad H|\omega
,m\rangle =\omega |\omega ,m\rangle , \\
O_{i}|\omega _{0},m\rangle &=&m_{i}|\omega _{0},m\rangle ,\quad O_{i}|\omega
,m\rangle =m_{i}|\omega ,m\rangle .
\end{eqnarray*}
As at the statistical quantum level we only measure mean values of
observables \footnote{%
Other quantum measurements imply a limit from category B to category A (cf. 
\cite{Deco}) which is outside the scope of this paper.} let us define the
observables we will use. The most general observable that we are going to
consider in our model reads: 
\begin{eqnarray}
O &=&\sum_{mm^{\prime }}O(\omega _{0})_{mm^{\prime }}|\omega _{0},m\rangle
\langle \omega _{0},m^{\prime }|+\sum_{mm^{\prime }}\int_{0}^{\infty
}d\omega O(\omega )_{mm^{\prime }}|\omega ,m\rangle \langle \omega
,m^{\prime }|+  \nonumber \\
&&+\sum_{mm^{\prime }}\int_{0}^{\infty }d\omega O(\omega ,\omega
_{0})_{mm^{\prime }}|\omega ,m\rangle \langle \omega _{0},m^{\prime }|+ 
\nonumber \\
&&+\sum_{mm^{\prime }}\int_{0}^{\infty }d\omega ^{\prime }O(\omega
_{0},\omega ^{\prime })_{mm^{\prime }}|\omega _{0},m\rangle \langle \omega
^{\prime },m^{\prime }|+  \nonumber \\
&&+\sum_{mm^{\prime }}\int_{0}^{\infty }\int_{0}^{\infty }d\omega d\omega
^{\prime }O(\omega ,\omega ^{\prime })_{mm^{\prime }}|\omega ,m\rangle
\langle \omega ^{\prime },m^{\prime }|,  \label{2.5}
\end{eqnarray}
where $O^{\dagger }=O$ and $O(\omega )_{mm^{\prime }}$, $O(\omega ,\omega
_{0})_{mm^{\prime }}$, $O(\omega _{0},\omega )_{mm^{\prime }}$ and $O(\omega
,\omega ^{\prime })_{mm^{\prime }}$ are ordinary functions of the real
variables $\omega $ and $\omega ^{\prime }$(these functions must have some
mathematical properties in order to develop the theory; these properties are
listed in paper \cite{CyLII}). We will say that these observables belong to
a space ${\cal O}$ (which is contained in the algebra studied in \cite{Ant}%
). This space has the {\it basis} $\{|\omega _{0},mm^{\prime })$, $|\omega
,mm^{\prime })$, $|\omega \omega _{0},mm^{\prime })$, $|\omega _{0}\omega
^{\prime },mm^{\prime })$, $|\omega \omega ^{\prime },mm^{\prime })\}$: 
\[
|\omega _{0},mm^{\prime })\doteq |\omega _{0},m\rangle \langle \omega
_{0},m^{\prime }|,\quad |\omega ,mm^{\prime })\doteq |\omega ,m\rangle
\langle \omega ,m^{\prime }|, 
\]
\begin{equation}
|\omega \omega _{0},mm^{\prime })\doteq |\omega ,m\rangle \langle \omega
_{0},m^{\prime }|,\quad |\omega _{0}\omega ^{\prime },mm^{\prime })\doteq
|\omega _{0},m\rangle \langle \omega ^{\prime },m^{\prime }|,  \label{2.5'}
\end{equation}
\[
|\omega \omega ^{\prime },mm^{\prime })\doteq |\omega ,m\rangle \langle
\omega ^{\prime },m^{\prime }| 
\]
The quantum states $\rho $ are measured by the observables just defined,
computing the mean values of these observable in the quantum states, i. e.
in the usual notation: $\langle O\rangle _{\rho }=Tr(\rho ^{\dagger }O).$ We
can consider that mean values are the more primitive objects of quantum
theory \cite{Ballentine}. These mean values, generalized as in paper \cite
{Fund}, can be considered as linear functionals $\rho $ (mapping the vectors 
$O$ on the real numbers)$,$ that we can call $(\rho |O)$ (see also \cite
{Bogo}). Then $\rho \in {\cal S\subset O}^{^{\prime }},$ where ${\cal S}$ is
a convenient convex set contained in ${\cal O}^{^{\prime }}$, the space of
linear functionals over ${\cal O}$ \cite{CyLI}, \cite{CyLIII}\footnote{%
We can choose more general ${\cal O}$ but the one we have defined is enough
for our purpose.}. The basis of ${\cal O}^{\prime }$ (that can also be
considered as the {\it co-basis} of ${\cal O)}$ is $\{(\omega
_{0},mm^{\prime }|$, $(\omega ,mm^{\prime }|$, $(\omega \omega
_{0},mm^{\prime }|$, $(\omega _{0}\omega ^{\prime },mm^{\prime }|$, $(\omega
\omega ^{\prime },mm^{\prime }|\}$ defined as functionals by the equations: 
\[
(\omega _{0},mm^{\prime }|\omega _{0},nn^{\prime })=\delta _{mn}\delta
_{m^{\prime }n^{\prime }},\quad (\omega ,mm^{\prime }|\eta ,nn^{\prime
})=\delta (\omega -\eta )\delta _{mn}\delta _{m^{\prime }n^{\prime }}, 
\]
\[
(\omega \omega _{0},mm^{\prime }|\eta \omega _{0},nn^{\prime })=\delta
(\omega -\eta )\delta _{mn}\delta _{m^{\prime }n^{\prime }}, 
\]
\[
(\omega _{0}\omega ^{\prime },mm^{\prime }|\omega _{0}\eta ^{\prime
},nn^{\prime })=\delta (\omega ^{\prime }-\eta ^{\prime })\delta _{mn}\delta
_{m^{\prime }n^{\prime }}, 
\]
\begin{equation}
(\omega \omega ^{\prime },mm^{\prime }|\eta \eta ^{\prime },nn^{\prime
})=\delta (\omega -\eta )\delta (\omega ^{\prime }-\eta ^{\prime })\delta
_{mn}\delta _{m^{\prime }n^{\prime }}.  \label{2.5''}
\end{equation}
and all other $(.|.)$ are zero. Then, a generic quantum state reads: 
\begin{eqnarray}
\rho &=&\sum_{mm^{\prime }}\overline{\rho (\omega _{0})}_{mm^{\prime
}}(\omega _{0},mm^{\prime }|+\sum_{mm^{\prime }}\int_{0}^{\infty }d\omega 
\overline{\rho (\omega )}_{mm^{\prime }}(\omega ,mm^{\prime }|+  \nonumber \\
&&+\sum_{mm^{\prime }}\int_{0}^{\infty }d\omega \overline{\rho (\omega
,\omega _{0})}_{mm^{\prime }}(\omega \omega _{0},mm^{\prime }|+  \nonumber \\
&&+\sum_{mm^{\prime }}\int_{0}^{\infty }d\omega ^{\prime }\overline{\rho
(\omega _{0},\omega ^{\prime })}_{mm^{\prime }}(\omega _{0}\omega ^{\prime
},mm^{\prime }|+  \nonumber \\
&&+\sum_{mm^{\prime }}\int_{0}^{\infty }d\omega \int_{0}^{\infty }d\omega
^{\prime }\overline{\rho (\omega ,\omega ^{\prime })}_{mm^{\prime }}(\omega
\omega ^{\prime },mm^{\prime }|  \label{2.6}
\end{eqnarray}
where 
\[
\overline{\rho (\omega ,\omega _{0})}_{mm^{\prime }}=\rho (\omega
_{0},\omega )_{m^{\prime }m},\quad \overline{\rho (\omega ,\omega ^{\prime })%
}_{mm^{\prime }}=\rho (\omega ^{\prime },\omega )_{m^{\prime }m}, 
\]
and $\overline{\rho (\omega _{0})}_{mm}$ and $\overline{\rho (\omega )}_{mm}$
are real and non negative satisfying the total probability condition 
\begin{equation}
(\rho |I)=\sum_{m}\rho (\omega _{0})_{mm}+\sum_{m}\int_{0}^{\infty }d\omega
\rho (\omega )_{mm}=1,  \label{2.6'}
\end{equation}
where $I=\sum_{m}|\omega _{0},m\rangle \langle \omega
_{0},m|+\int_{0}^{\infty }d\omega \sum_{m}|\omega ,m\rangle \langle \omega
,m|$ is the identity operator in ${\cal O}$. Eq. (\ref{2.6'}) is the
extension to state functionals of the usual condition $Tr\rho ^{\dagger }=1$%
, used when $\rho $ is a density operator. Thus, from now on, $Tr\rho \doteq
(\rho |I).$

The time evolution of the quantum state $\rho $ reads: 
\begin{eqnarray}
\rho (t) &=&\sum_{mm^{\prime }}\overline{\rho (\omega _0)}_{mm^{\prime
}}(\omega _0,mm^{\prime }|+\sum_{mm^{\prime }}\int_0^\infty d\omega 
\overline{\rho (\omega )}_{mm^{\prime }}(\omega ,mm^{\prime }|+  \nonumber \\
&&+\sum_{mm^{\prime }}\int_0^\infty d\omega \overline{\rho (\omega ,\omega
_0)}_{mm^{\prime }}e^{i(\omega -\omega _0)t}(\omega \omega _0,mm^{\prime }|+
\nonumber \\
&&+\sum_{mm^{\prime }}\int_0^\infty d\omega ^{\prime }\overline{\rho (\omega
_0,\omega ^{\prime })}_{mm^{\prime }}e^{i(\omega _0-\omega ^{\prime
})t}(\omega _0\omega ^{\prime },mm^{\prime }|+  \nonumber \\
&&+\sum_{mm^{\prime }}\int_0^\infty d\omega \int_0^\infty d\omega ^{\prime }%
\overline{\rho (\omega ,\omega ^{\prime })}_{mm^{\prime }}e^{i(\omega
-\omega ^{\prime })t}(\omega \omega ^{\prime },mm^{\prime }|  \label{2.7}
\end{eqnarray}

As we have already said at the statistical quantum level we can only measure
mean values of observables in quantum states, i. e.: 
\begin{eqnarray}
\langle O\rangle _{\rho (t)} &=&(\rho (t)|O)=  \nonumber \\
&=&\sum_{mm^{\prime }}\overline{\rho (\omega _{0})}_{mm^{\prime }}O(\omega
_{0})_{mm^{\prime }}+\sum_{mm^{\prime }}\int_{0}^{\infty }d\omega \overline{%
\rho (\omega )}_{mm^{\prime }}O(\omega )_{mm^{\prime }}+  \nonumber \\
&&+\sum_{mm^{\prime }}\int_{0}^{\infty }d\omega \overline{\rho (\omega
,\omega _{0})}_{mm^{\prime }}e^{i(\omega -\omega _{0})t}O(\omega ,\omega
_{0})_{mm^{\prime }}+  \nonumber \\
&&+\sum_{mm^{\prime }}\int_{0}^{\infty }d\omega ^{\prime }\overline{\rho
(\omega _{0},\omega ^{\prime })}_{mm^{\prime }}e^{i(\omega _{0}-\omega
^{\prime })t}O(\omega _{0},\omega ^{\prime })_{mm^{\prime }}+  \nonumber \\
&&+\sum_{mm^{\prime }}\int_{0}^{\infty }d\omega \int_{0}^{\infty }d\omega
^{\prime }\overline{\rho (\omega ,\omega ^{\prime })}_{mm^{\prime
}}e^{i(\omega -\omega ^{\prime })t}O(\omega ,\omega ^{\prime })_{mm^{\prime
}},  \label{2.8}
\end{eqnarray}
Let us now consider the fate of these mean values when $t\rightarrow \infty
, $ using the Riemann-Lebesgue theorem we obtain the limit, for all $O\in 
{\cal O}$ 
\begin{equation}
\lim_{t\rightarrow \infty }\langle O\rangle _{\rho (t)}=\langle O\rangle
_{\rho _{*}}  \label{2.9}
\end{equation}
where we have introduced the diagonal asymptotic or equilibrium state
functional\footnote{%
The three of diagonal terms of eqs. (\ref{2.7}) or (\ref{2.8}) disappear.
This happens because we have just one bound state and would not be the case
for more than one bound state.} 
\begin{equation}
\rho _{*}=\sum_{mm^{\prime }}\overline{\rho (\omega _{0})}_{mm^{\prime
}}(\omega _{0},mm^{\prime }|+\sum_{mm^{\prime }}\int_{0}^{\infty }d\omega 
\overline{\rho (\omega )}_{mm^{\prime }}(\omega ,mm^{\prime }|  \label{2.10}
\end{equation}
Therefore, in a weak sense we have: 
\begin{equation}
W\lim_{t\rightarrow \infty }\rho (t)=\rho _{*}  \label{2.11}
\end{equation}
Thus, any quantum state weakly goes to a linear combination of the energy
diagonal states $(\omega _{0},mm^{\prime }|$ and $(\omega ,mm^{\prime }|$
(the energy ''off-diagonal'' states $(\omega \omega _{0},mm^{\prime }|$, $%
(\omega _{0}\omega ^{\prime },mm^{\prime }|$ and $(\omega \omega ^{\prime
},mm^{\prime }|$ are not present in $\rho _{*}$). This is the case if we
observe and measure the system evolution with {\it any possible observable
of space }${\cal O}$ (albeit the discussion in section IV.A.3){\it .} Then,
from the observational point of view, we have decoherence of the energy
levels, even that, from the strong limit point of view the off-diagonal
terms never vanish, they just oscillate, since we cannot directly use the
Riemann-Lebesgue theorem in the operator equation (\ref{2.7}).

\subsubsection{Decoherence in the other ''momentum'' dynamical variables.}

Having established the decoherence in the energy levels we must consider the
decoherence in the other dynamical variables $O_i$, of the CSCO where we are
working. We will call these variables ''momentum variables''. As the
expression of $\rho _{*}$ given in eq. (\ref{2.10}) involve only the time
independent components of $\rho (t)$, it is impossible that a different
decoherence process take place to eliminate the off-diagonal terms in the
remaining $N$ dynamical variables. Therefore, the only thing to do is to
find if there is a basis where the off-diagonal components of $\rho (\omega
_0)_{mm^{\prime }}$ and $\rho (\omega )_{mm^{\prime }}$ vanish at any time.

Let us consider the following change of basis 
\begin{equation}
|\omega _0,r\rangle =\sum_mU(\omega _0)_{mr}|\omega _0,m\rangle ,\qquad
|\omega ,r\rangle =\sum_mU(\omega )_{mr}|\omega ,m\rangle ,  \label{2.11'}
\end{equation}
where $r$ and $m$ are short notations for $r\doteq \{r_1,...,r_N\}$ and $%
m\doteq \{m_1,...,m_N\}$, and $\left[ U(\Omega )^{-1}\right] _{mr}=\overline{%
U(\Omega )}_{rm}$ ($\Omega $ denotes either $\omega _0<0$ or $\omega \in 
{\Bbb R}^{+}$).

The new basis $\{|\omega _{0},r\rangle ,|\omega ,r\rangle \}$ verifies the
generalized orthogonality conditions 
\begin{eqnarray*}
\langle \omega _{0},r|\omega _{0},r^{\prime }\rangle &=&\delta _{rr^{\prime
}},\quad \langle \omega ,r|\omega ^{\prime },r^{\prime }\rangle =\delta
(\omega -\omega ^{\prime })\delta _{rr^{\prime }}, \\
\langle \omega _{0},r|\omega ,r^{\prime }\rangle &=&\langle \omega ,r|\omega
_{0},r^{\prime }\rangle =0.
\end{eqnarray*}

As $\overline{\rho (\omega _0)}_{mm^{\prime }}=\rho (\omega _0)_{m^{\prime
}m}$ and $\overline{\rho (\omega )}_{mm^{\prime }}=\rho (\omega )_{m^{\prime
}m}$, it is possible to choose $U(\omega _0)$ and $U(\omega )$ in such a way
that the off-diagonal parts of $\rho (\omega _0)_{rr^{\prime }}$ and $\rho
(\omega )_{rr^{\prime }}$ vanish, i.e. 
\begin{equation}
\rho (\omega _0)_{rr^{\prime }}=\rho _r(\omega _0)\,\delta _{rr^{\prime
}},\qquad \rho (\omega )_{rr^{\prime }}=\rho _r(\omega )\,\delta
_{rr^{\prime }}.  \label{nonint}
\end{equation}
Therefore, there is a{\it \ final pointer basis} for the observables given
by $\{|\omega _0,rr^{\prime })$, $|\omega ,rr^{\prime })$, $|\omega \omega
_0,rr^{\prime })$, $|\omega _0\omega ^{\prime },rr^{\prime })$, $|\omega
\omega ^{\prime },rr^{\prime })\}$ and defined as in eq. (\ref{2.5'}). The
corresponding final pointer basis for the states $\{(\omega _0,rr^{\prime }|$%
, $(\omega ,rr^{\prime }|$, $(\omega \omega _0,rr^{\prime }|$, $(\omega
_0\omega ^{\prime },rr^{\prime }|$, $(\omega \omega ^{\prime },rr^{\prime
}|\}$ diagonalizes the time independent part of $\rho (t)$ and therefore it
diagonalizes the final state $\rho _{*}$%
\begin{equation}
\rho _{*}=W\lim_{t\rightarrow \infty }\rho (t)=\sum_r\rho _r(\omega
_0)(\omega _0,rr|+\sum_r\int_0^\infty d\omega \rho _r(\omega )(\omega ,rr|.
\label{RO1}
\end{equation}

Now we can define the{\it \ final exact pointer observables} \cite{Zurek}%
\footnote{%
If we would like to just have integrals and treat all spectra as continuous
we would write 
\[
\sum_{r}\int_{0}^{\infty }d\omega =\int_{0}^{\infty }...\int_{0}^{\infty
}d\omega dR_{1}...dR_{N}\delta (r_{1}-R_{1})...\delta (r_{N}-R_{N})=
\]
\begin{equation}
\int d\omega \int d\mu (R)  \label{fn}
\end{equation}
} 
\begin{equation}
P_{i}=\sum_{r}P_{r}^{i}(\omega _{0})|\omega _{0},r\rangle \langle \omega
_{0},r|+\int_{0}^{\infty }d\omega \sum_{r}P_{r}^{i}(\omega )|\omega
,r\rangle \langle \omega ,r|.  \label{RO2}
\end{equation}
As $H$ and $P_{i}$ are diagonal in the basis $\{|\omega _{0},r\rangle $, $%
|\omega ,r\rangle \}$, the set $\{H,P_{i},...P_{N}\}$ is precisely the
complete set of commuting observables (CSCO) \footnote{%
In usual quantum mechanical system this set is numerable.} related to this
basis, where $\rho _{*}$ is diagonal in the corresponding co-basis for the
states. For simplicity we define the operators $P_{i}$ such that $%
P_{r}^{i}(\omega _{0})=P_{r}^{i}(\omega )=r_{i}$, thus 
\begin{equation}
P_{i}|\omega _{0},r\rangle =r_{i}|\omega _{0},r\rangle ,\qquad P_{i}|\omega
,r\rangle =r_{i}|\omega ,r\rangle .  \label{RO3}
\end{equation}
Therefore $\{|\omega _{0},r\rangle $, $|\omega ,r\rangle \}$ is the
observers' final pointer basis were there is a perfect decoherence in the
corresponding state co-basis. Moreover the generalized states $(\omega
_{0},rr|$ and $(\omega ,rr|$ are constants of the motion, and therefore
these exact pointer observables have a constant statistical entropy and will
be ''at the top of the list'' of Zurek's ''predictability sieve'' \cite
{Zurek}.

Therefore:

i.- Decoherence in the energy is produced by the time evolution.

ii.- Decoherence in the other dynamical variables can be seen if we choose
an adequate basis, namely the final pointer basis.

Our main result is eq. (\ref{RO1}): {\it When }$t\rightarrow \infty $ {\it %
then }$\rho (t)\rightarrow \rho _{*}$ {\it and in this state the dynamical
variables }$H,P_{1},...,P_{N}$ {\it are well defined. Therefore the eventual
conjugated variables to these momentum variables (namely: configuration
variables, if they exist) are completely undefined.}

In fact, calling by ${\Bbb L}_{i}$ the generator of the displacements along
the eventual configuration variable conjugated to $P_{i}$, we have $({\Bbb L}%
_{i}\rho _{*}|O)=(\rho _{*}|{\Bbb L}_{i}^{\dagger }O)=(\rho
_{*}|[P_{i},O])=0 $ for all $O\in {\cal O}$ as it can be proved by direct
computation \cite{Deco}. Then $\rho _{*}$\ is homogeneous in these
configuration variables.

From the preceding section we may have the feeling that the process of
decoherence must be found in all the physical systems, and therefore, all of
them eventually would become classical when $\hbar \rightarrow 0$. It is not
so, this evolution only happens under certain conditions, as explained in 
\cite{Deco}. I. e., if there is more than one bound state or the decoherence
time is infinite the system does not decohere.

\subsection{Quantum thermalization.}

\subsubsection{The minimally biased state.}

At this point we can find the final quantum minimally biased state according
to information theory \cite{Jaynes}. For simplicity let us just use one
observable in our CSCO, the hamiltonian $H$ with no bound state $\omega _{0}$%
\footnote{%
In fact, in our model $H$ is the only member of the CSCO with continuous
spectrum, so all the problem is contained in $H.$}. Then eq. (\ref{RO1})
reads 
\begin{equation}
(\rho _{*}|=\int_{0}^{\infty }\rho (\omega )(\omega |d\omega  \label{J.1}
\end{equation}
$(\rho _{*}|$ has trace one, so from eq. (\ref{2.6'}) we know that 
\begin{equation}
Tr\rho _{*}=(\rho _{*}|I)=\int_{0}^{\infty }\rho (\omega )d\omega =1
\label{J.2}
\end{equation}
The mean value of the energy reads 
\begin{equation}
\langle H\rangle _{\rho _{*}}=(\rho _{*}|H)=\int_{0}^{\infty }\omega \rho
(\omega )d\omega =E  \label{J.3}
\end{equation}
If this is the only data available to obtain the minimal biased state we
must maximizes the (Shannon) missing information \footnote{%
I. e., the continuous version of eq. (2.3) or \cite{Jaynes} or eq. (5.6) of 
\cite{Katz}, postulating that when no information or probabilities are
available we should consider the cells equally probable.} 
\begin{equation}
{\cal I=H}[\rho _{*}]=-K\int_{0}^{\infty }\rho (\omega )\log \rho (\omega
)d\omega  \label{J.4}
\end{equation}
contrained by eqs. (\ref{J.2}) and (\ref{J.3}). The solution of this
variational problem is $\rho (\omega )\sim e^{-\beta \omega }$ so the
unbiased $(\rho _{*}|$ is 
\begin{equation}
(\rho _{*}|=Z^{-1}\int_{0}^{\infty }e^{-\beta \omega }(\omega |d\omega
\label{J.5}
\end{equation}
where the constants $Z$ and $\beta $ can be computed from eqs. (\ref{J.2})
and (\ref{J.3}).

The next problem is to represent this minimal biased $(\rho _{*}|$ as $%
e^{-\beta H}.$ But an equation like $(\rho _{*}|$=$e^{-\beta H}$ is
impossible since $H$ is a ''vector'' (ket) and $(\rho _{*}|$ is a
''functional'' (bra). To solve this problem we must go back to eq. (\ref{J.2}%
) and see that the trace of the state can be written as 
\begin{equation}
Tr\rho _{*}=(\rho |I)=(\rho |\int_{0}^{\infty }|\omega )d\omega ;\qquad
|I)=\int_{0}^{\infty }|\omega )d\omega  \label{J.6}
\end{equation}
Symmetrically we can define the trace of an operator $A$ as

\begin{equation}
TrA=(I|A)=\int_{0}^{\infty }(\omega |d\omega |A);\qquad (I|=\int_{0}^{\infty
}(\omega |d\omega  \label{J.7}
\end{equation}
We will see that this definition is completely reasonable at the end of
section IV.B. Using it we can write the functional that corresponds to $%
(\rho _{*}|$ as 
\begin{equation}
w_{\beta }[A]=(\rho _{*}|A)=\frac{(I|e^{-\beta H}A)}{(I|e^{-\beta H})}
\label{J.8}
\end{equation}
We can immediately see that the trace of this functional is one since $%
w_{\beta }[I]=(\rho _{*}|I)=1.$ Moreover, the last definition coincides with
the one of eq. (\ref{J.5}) since $e^{-\beta H}=\int e^{-\beta \omega
}|\omega )d\omega $ and 
\begin{equation}
Z=(I|e^{-\beta H})=\int \int (\omega |e^{-\beta \omega }|\omega ^{\prime
})d\omega d\omega ^{\prime }=\int e^{-\beta \omega }\delta (\omega -\omega
^{\prime })d\omega d\omega ^{\prime }=\int e^{-\beta \omega }d\omega
\label{J.9}
\end{equation}
Also 
\begin{equation}
(I|e^{-\beta H}A)=\int e^{-\beta \omega }(\omega |A)d\omega  \label{J.10}
\end{equation}
From the two last equations it is evident that (\ref{J.5}) coincides with (%
\ref{J.8}) and we can write the minimally biased canonical final state as 
\begin{equation}
(\rho _{*}|=Z^{-1}(I|e^{-\beta H}  \label{J.11}
\end{equation}
It is easy to generalize the above reasoning from $H$ to the CSCO $%
\{H,P_{i},...P_{N}\}.$ Then we would obtain 
\begin{equation}
(\rho _{*}|=Z^{-1}(I|e^{-\beta H-\gamma _{1}P_{1}-...-\gamma _{N}P_{N}}
\label{J.11'}
\end{equation}
that corresponds to a generalized grand-canonical ensemble.

\subsubsection{KMS equilibrium.}

As a demonstration that $w_{\beta }[A]$ is the good equilibrium limit at
temperature $\beta ^{-1}$ we will prove that it satisfies the KMS condition 
\cite{Haag}. The time evolution of operator $A$ is given by 
\begin{equation}
\alpha _{t}(A)=e^{iHt}Ae^{-iHt}  \label{J.12}
\end{equation}
( eq. (\ref{2.7}), for the state evolution, is deduced from this equation).
Then we must check the analyticity properties of 
\begin{equation}
w_{\beta }[\alpha _{t}(A)B]=\frac{(I|e^{-\beta H}\alpha _{t}(A)B)}{%
(I|e^{-\beta H})}=\frac{(I|e^{-\beta H}e^{iHt}Ae^{-iHt}B)}{(I|e^{-\beta H})}
\label{J.13}
\end{equation}
We must first prove that definition (\ref{J.7}) has the cyclic trace
property. Since the algebra ${\cal O}$ is associative \cite{Ant} we only
need to demonstrate the commutativity. In fact\footnote{%
All these probabilities naturally appear if we postulate that the
characteristic algebra is a nuclear one as we will do in the rigorous
treatment of the subject elsewhere.} 
\begin{equation}
Tr(AB)=(I|AB)=(\int (\omega |d\omega |AB)=\int A_{\omega }B_{\omega }d\omega
=(I|BA)=Tr(BA)  \label{J.14}
\end{equation}
then 
\begin{equation}
w_{\beta }[\alpha _{t}(A)B]=\frac{(I|Be^{iH(t+i\beta )}Ae^{-iHt})}{%
(I|e^{-\beta H})}=\frac{(I|Be^{-\beta H}e^{-\beta H}e^{iHt}Ae^{-iHt}e^{\beta
H})}{(I|e^{-\beta H})}=w_{\beta }(B\alpha _{t+i\beta }(A))  \label{J.15}
\end{equation}
Then calling (\cite{Haag} eq. (V.1.7)) 
\[
F_{A,B}^{(\beta )}(z)=w_{\beta }[B\alpha _{z}(A)] 
\]
\begin{equation}
G_{A,B}^{(\beta )}(z)=w_{\beta }[\alpha _{z}(A)B]  \label{J.16}
\end{equation}
where $z=t+i\gamma ,$ we obtain 
\begin{equation}
F_{A,B}^{(\beta )}(z)=\frac{(I|e^{-\beta H}Be^{iHz}Ae^{-iHz})}{(I|e^{-\beta
H})}=\frac{(I|Be^{iHt}e^{-\gamma H}Ae^{-iHt}e^{-(\beta -\gamma )H})}{%
(I|e^{-\beta H})}  \label{J.17}
\end{equation}
But from eq., (\ref{J.7}) 
\begin{equation}
Tre^{-\beta H}=(I|e^{-\beta H})=\int e^{-\beta \omega }d\omega  \label{J.18}
\end{equation}
which is only convergent and for $\beta >0.$ Analogously, the r.h.s. of eq. (%
\ref{J.17}) reads 
\[
\int \int B(\omega ,\omega ^{\prime })e^{i\omega ^{\prime }t}e^{-\gamma
\omega ^{\prime }}A(\omega ^{\prime },\omega )e^{-i\omega t}e^{-(\beta
-\gamma )\omega }d\omega d\omega ^{\prime }/\int e^{-\beta \omega }d\omega 
\]
Thus in order that $F_{A,B}^{(\beta )}(z)$ be analytic in $z$ these
integrals must be convergent and therefore the three real parts of the
exponents: $\beta ,$ $\gamma $ and, $\beta -\gamma $ must be positive. Then
the operator in (\ref{J.17}) is analytic for 
\begin{equation}
0<\gamma <\beta  \label{J.19}
\end{equation}
Also from (\ref{J.15}) 
\begin{equation}
G_{A,B}^{(\beta )}(t)=F_{A,B}^{(\beta )}(t+i\beta )  \label{J.20}
\end{equation}
and both functions are analytic for the $z=t+i\gamma $ satisfying condition (%
\ref{J.19}). Therefore KMS condition is satisfied.

We have proved that the thermal equilibrium state exists in the system but
we have not proved if the system spontaneously goes to this equilibrium (it
''thermalizes'') and we have not said under what conditions this phenomenon
takes place. The reason is that the two powerful devises we have used: the
diagonal form of hamiltonian $H$ and information theory allow us to reach
the above conclusion but they yield to a static and already thermalized
diagonal of the density matrix. To study thermalization we must consider at
least two subsystems and their interaction {\it before}\footnote{%
Or after diagonalization but in an indirect way as in \cite{Katz} or \cite
{L&L}} diagonalization and show how they reach the same temperature. In this
case the method can be also used to show that, if the system is endowed with
an adequate interaction, it naturally evolves to state (\ref{J.11}), i. e.
it ''thermalizes''. Thus the kind of interaction defines if the system
thermalize or not. In fact, we have proved in paper \cite{Fund} that the
thermodynamic limit can by formalized with our method and we have developed
a simple model with linear interaction (Friedrichs model) where an
oscillator is thermalized by a bath for small interaction. The case of big
interactions will be treated elsewhere.

Let us finally remark that a KMS state has a trivial explicit definition in
a box. But in an unbounded space it was substituted by a state that
satisfies KMS conditions with no explicit expression \cite{Haag}. Our method
has allow us to find this explicit expression, namely (\ref{J.11}).

\section{The classical statistical limit.}

In this section we will use the Wigner integrals that introduce an
isomorphism between quantum observables $O$ and states $\rho $ and their
classical analogues $O^W(q,p)$ and $\rho ^W(q,p)$ \cite{Wigner}: 
\begin{eqnarray}
O^W(q,p) &=&\int d\lambda \,\langle q-\frac \lambda 2|O|q+\frac \lambda 2%
\rangle \,\exp (i\lambda p)  \nonumber \\
\rho ^W(q,p) &=&\frac 1{\pi ^{N+1}}\int d\lambda \,(\rho ||q+\lambda \rangle
\langle q-\lambda |)\,\exp (2i\lambda p).  \label{Wig}
\end{eqnarray}

It is possible to prove that $\int dq\,dp\,\rho ^{W}(q,p)=(\rho |I)=1$, but
in general $\rho ^{W}$ is not always non negative. It is also possible to
deduce that 
\begin{equation}
(\rho ^{W}|O^{W})=\int dq\,dp\,\rho ^{W}(q,p)O^{W}(q,p)=(\rho |O),
\label{mean}
\end{equation}
and therefore: to the mean value in the classical space corresponds the mean
value in the quantum space. Moreover, calling $L$ the classical Liouville
operator, and ${\Bbb L}$ the quantum Liouville-Von Neumann operator, we have 
\begin{equation}
L\left[ \rho ^{W}(q,p)\right] =\left[ {\Bbb L}\rho \right] ^{W}(q,p)+O(\hbar
/s),  \label{ara1}
\end{equation}
where $L\,\rho ^{W}(q,p)=i\left\{ H^{W}(q,p),\rho ^{W}(q,p)\right\} _{PB},$ $%
s$ is an action with the characteristic dimension of the system, and 
\begin{equation}
({\Bbb L}\rho |O)=(\rho |[H,O]).  \label{ara2}
\end{equation}
Finally, if $O=O_{1}O_{2}$, where $O_{1}$ and $O_{2}$ are two quantum
observables, we have 
\begin{equation}
O^{W}(q,p)=O_{1}^{W}(q,p)O_{2}^{W}(q,p)+O(\hbar /s).  \label{pro}
\end{equation}

We will prove that the distribution function $\rho _{*}^{W}(q,p)$, that
corresponds to the state functional $\rho _{*}$ via the Wigner integral is a
non negative function of the classical constants of the motion, in our case $%
\footnote{%
We can prove that $H^{W}(q,p)$, $P_{1}^{W}(q,p)$,..., $P_{N}^{W}(q,p)$ are
constants of the motion using Heisenberg version of Liouville equation and
eq. (\ref{ara1}).},$ obtained from the corresponding quantum operators $H$, $%
P_{1}$,..., $P_{N}$.

From eq. (\ref{RO1}) we have: 
\begin{equation}
\rho _{*}=W\lim_{t\rightarrow \infty }\rho (t)=\sum_{r}\rho _{r}(\omega
_{0})(\omega _{0},rr|+\sum_{r}\int_{0}^{\infty }d\omega \rho _{r}(\omega
)(\omega ,rr|,  \label{5.2}
\end{equation}
so we must compute: 
\begin{equation}
\rho _{\omega r}^{W}(q,p)\doteq \pi ^{-N-1}\int (\omega ,rr||q+\lambda
\rangle \langle q-\lambda |)e^{2ip\lambda }d\lambda   \label{5.3}
\end{equation}
We know from \cite{Fund} section II. C, (or we can prove directly from eqs.(%
\ref{RO1}-\ref{RO3})) that 
\begin{eqnarray}
(\omega _{0},rr|H^{n}) &=&\omega _{0}^{n},\quad (\omega ,rr|H^{n})=\omega
^{n},  \nonumber \\
(\omega _{0},rr|P_{i}^{n}) &=&r_{i}^{n},\quad (\omega
,rr|P_{i}^{n})=r_{i}^{n},\quad i=1,...,N  \label{5.4}
\end{eqnarray}
for $n=0,1,2,...$ Using the relation (\ref{pro}) between quantum and
classical products of observables and relation (\ref{mean}) between quantum
and classical mean values, in the limit $\hbar \rightarrow 0$ (precisely
when $s\rightarrow \infty ,$ i. e. when the dimension of the system are very
large compared with $\hbar )$\footnote{%
We will consider that we always take this limit when we refer to classical
equations below.} we deduce that the characteristic property of the
distribution $\rho _{\omega r}^{W}(q,p)$, that corresponds to the state
functional $(\omega ,rr|$, is\footnote{%
To simplify the demonstration we will consider that all the spectra are
continuous as explained in footnote 7.}: 
\begin{equation}
\int \rho _{\omega r}^{W}(q,p)[H^{W}(q,p)]^{n}dqdp=\omega ^{n},\quad \int
\rho _{\omega r}^{W}(q,p)[P_{i}^{W}(q,p)]^{n}dqdp=r_{i}^{n},  \label{5.5}
\end{equation}
for any natural number $n.$ Thus $\rho _{\omega r}^{W}(q,p)$ must be the
functional\footnote{%
Let us precisely define the $\delta $ of the next equation$.$ In eq. (\ref
{5.5}$_{2}$) we can make the canonical transformation $q,p\rightarrow
Q^{W},P^{W}$ then it reads 
\[
\int \rho _{\omega
r}^{W}(Q_{i}^{W},P_{i}^{W})[P_{i}^{W}]^{n}dQ_{i}^{W}dP_{i}^{W}=r_{i}^{n},%
\qquad i=0.1,...N
\]
Then: 
\[
\rho _{\omega r}^{W}(Q^{W},P^{W})=(V_{Q})^{-1}\prod_{i=0}^{N}\delta
(P_{i}^{W}-r_{i})
\]
where 
\[
V_{Q}=\int dQ^{W}
\]
is the volume of configuration space that we will consider as bounded for
simplicity. Nevertheless the case $V_{Q}\rightarrow \infty $ can be studied
with the techniques of paper \cite{Fund}.
\par
Then to prove eqs. (\ref{5.6}) and (\ref{5.6'}) we can first write eq. (\ref
{5.2}) as 
\[
\rho _{*}^{W}(q,p)=\int d\mu (R)\rho _{R}(\omega _{0})\rho _{\omega
_{0}R}^{W}(q,p)+
\]
\[
\int d\mu (R)\int_{0}^{\infty }d\omega \rho _{R}(\omega )\rho _{\omega
R}^{W}(q,p)
\]
where we have used eq. (\ref{fn}).} 
\begin{equation}
\rho _{\omega r}^{W}(q,p)=\delta (H^{W}(q,p)-\omega )\delta
(P_{1}^{W}(q,p)-r_{1})...\delta (P_{N}^{W}(q,p)-r_{N}).  \label{5.6}
\end{equation}
For the distribution $\rho _{\omega _{0}r}^{W}(q,p)$ corresponding to the
state functional $(\omega _{0},rr|$, we obtain 
\begin{equation}
\rho _{\omega _{0}r}^{W}(q,p)=\delta (H^{W}(q,p)-\omega _{0})\delta
(P_{1}^{W}(q,p)-r_{1})...\delta (P_{N}^{W}(q,p)-r_{N}).  \label{5.6'}
\end{equation}
Therefore, going back to eq. (\ref{5.2}) and since the Wigner relation is
linear, we have\footnote{%
We will call: 
\[
\rho _0(\omega ,r_1,...,r_N)=\rho _r(\omega _0)=\rho _{r_1,...,r_N}(\omega
_0) 
\]
\par
\[
\rho (\omega ,r_{1},...,r_{N})=\rho _{r}(\omega )=\rho
_{r_{1},...,r_{N}}(\omega )
\]
}: 
\begin{equation}
\rho _{*}^{W}(q,p)=\sum_{r}\rho _{r}(\omega _{0})\rho _{\omega
_{0}r}^{W}(q,p)+\sum_{r}\int_{0}^{\infty }d\omega \rho _{r}(\omega )\rho
_{\omega r}^{W}(q,p)=  \label{5.7}
\end{equation}
\[
\rho _{0}(H^{W}(q,p),P_{1}^{W}(q,p),...,P_{N}^{W}(q,p))+\rho
(H^{W}(q,p),P_{1}^{W}(q,p),...,P_{N}^{W}(q,p))
\]
Also we obtain $\rho _{*}^{W}(q,p)\geq 0$, because $\rho _{r}(\omega _{0})$
and $\rho _{r}(\omega )$ are non negative.

Therefore, the classical state $\rho _{*}^{W}(q,p)$ is a linear combination
of the generalized classical states $\rho _{\Omega r}^{W}(q,p)$ (where $%
\Omega $ is either $\omega _{0}$ or $\omega $), having well defined values $%
\Omega $, $r_{1}$,..., $r_{N}$ of the classical observables $H^{W}(q,p)$, $%
P_{1}^{W}(q,p)$,..., $P_{N}^{W}(q,p)$. The corresponding classical
canonically conjugated variables are completely undefined since neither $%
\rho _{\Omega r}^{W}(q,p)$ nor $\rho _{*}^{W}(q,p)$ is a function of these
variables. {\it So we reach, in the classical case, to the same conclusion
than in the quantum case }(see end of subsection II.A.2){\it . }But now all
the classical canonically conjugated variables $a_{0},a_{1},...,a_{N}$ do
exist since they can be found solving the corresponding Poisson brackets
differential equations. We can also expand the densities given in eqs. (\ref
{5.6}-\ref{5.7}) in terms of classical motions as shown in \cite{Deco} in
great detail. In fact, as the momenta $H^{W},P_{1}^{W},...,P_{N}^{W}$, or
any function of these momenta, that we will call generically $\Pi ,$ are
also constant of the motion, then we have $\frac{d}{dt}\Pi =-\partial
H/\partial \alpha =0$, where $\alpha $ is the classically conjugated
variable to $\Pi .$ So $H$ is just a function of the $\Pi $ and: 
\begin{equation}
\frac{d}{dt}\alpha =\frac{\partial H(\Pi )}{\partial \Pi }=\varpi (\Pi
)=const.  \label{5.9}
\end{equation}
so: 
\begin{equation}
\alpha _{j}(t)=\varpi _{j}(\Pi )t+\alpha _{j}(0),\qquad j=0,1,...,N.
\label{5.10}
\end{equation}
Thus (going back to the old coordinates $q,p$) in the set of classical
motions contained in the densities (\ref{5.6}) and (\ref{5.6'}) the momenta $%
H^{W}(q,p)$, $P_{1}^{W}(q,p)$,..., $P_{N}^{W}(q,p)$, are completely defined
and the origin of the corresponding motions, that we will respectively call $%
a_{0}(0)$, $a_{1}(0)$,...and $a_{N}(0)$, are completely undefined (since $%
\rho _{*}^{W}(q,p)$ does not contain these variables), in such a way that
the motions represented in the last equation homogeneously fill the surface
where $H^{W}$, $P_{1}^{W}$,..., and $P_{N}^{W}$, have constant values. If
the system is bounded and integrable (see below), these surface turns out to
be the a usual torus of phase space. This is the interpretation that we give
to the density (\ref{5.7}) which is just a function of the variables $H^{W}$%
, $P_{1}^{W}$,..., $P_{N}^{W},$ but it is not a function of the classical
conjugated variables $a_{0}$, $a_{1}$,..., $a_{N}$. The classical motions
described by eqs. (\ref{5.9}) and (\ref{5.10}) will be used to characterize
the dynamical system in the next section. Precisely, using these motions the
evolution of the $\rho ^{W},$ close to equilibrium $\rho _{*}^{W}$ can be
interpreted as Frobenius-Perron operators\cite{Mackey}\footnote{%
This is an important step towards classicality since the system has lost its
quantum characteristic (e. g. totality, contextuallity and non-locality
according to \cite{D. Bhom}) and we can resolve the classical density matrix
evolution in a set of particle motions.}.

In conclusion.

i.- We have shown that the quantum state functional $\rho (t)$ evolves to a
diagonal state $\rho _{*}$.

ii.- This quantum state $\rho _{*}$ has its corresponding classical density $%
\rho _{*}^W(q,p)$.

iii.- This classical density can be decomposed in classical densities that
correspond to sets of classical motions where $H^W$, $P_1^W$,..., $P_N^W$
remain constant. These motions have $H^W(q,p)$, $P_1^W(q,p)$,..., $%
P_N^W(q,p)=const$ and initial conditions $a_0(0),$ $a_1(0),$ $...,$ $a_N(0)$
distributed in an homogeneous way.

iv.- From eqs. (\ref{5.6}-\ref{5.7}) we will obtain that\footnote{%
From now on we will forget the bound eigenvalue $\omega _0.$} 
\begin{equation}
\rho _{*}^W(q,p)=\rho (H^W(q,p),P_1^W(q,p),...,P_N^W(q,p))\geq 0
\label{otra}
\end{equation}

\section{Relation between Classical Statistic, Dynamical Systems and
Thermodynamics .}

In this section we will characterize the classical motions we have obtained
in the ergodic hierarchy and we will make contact with classical
thermodynamics.

\subsection{The microcanonical ensemble.}

\subsubsection{The notion of isolating and non-isolating constant of the
motion.}

The classical constants of the motion $H^W(q,p),$ $P_1^W(q,p),...,$ $%
P_N^W(q,p)$ can be rigorously classified as \cite{Arnold}, \cite{Tabor}, 
\cite{Balescu}:

i.- {\bf Global or isolating constant of the motion, }that we will call $"H"$%
, precisely: 
\begin{equation}
H_{0}^{W}(q,p)=H^{W}(q,p),\text{ }%
H_{1}^{W}(q,p)=P_{1}^{W}(q,p),...,H_{A}^{W}(q,p)=P_{A}^{W}(q,p)  \label{3.1}
\end{equation}
when the conditions 
\begin{equation}
H_{i}^{W}=r_{i},\qquad i=0,...,A  \label{3.3}
\end{equation}
(where $r_{0}=\Omega )$ define global sub manifolds (tori in the bounded
case) ${\cal M}(r_{0},...,r_{A}),$ of phase space where the trajectories
necessarily move, for each set of constants $(r_{0},...,r_{A})$\footnote{%
Isolating constant of the motion would be the ''simple'' constant of the
motion in \cite{Katz} p. 60.}. The dimension of ${\cal M}$ is $%
2(N+1)-(A+1)=2N$ $-A+1$

{\bf ii.- Local or non isolating constant of the motion, }that we will call: 
\begin{equation}
J_{1}^{W}(q,p)=P_{A+1}^{W}(q,p),...,J_{N-A}^{W}(q,p)=P_{N}^{W}(q,p) 
\nonumber
\end{equation}
when the conditions 
\begin{equation}
J_{j}^{W}=r_{j+A}\qquad j=1,..,N-A  \label{3.4}
\end{equation}
do {\it not} define any global sub-manifold, since they are just local 
\footnote{%
Let us list the introduced dimension:
\par
i.-The total dimension is $2(N+1).$%
\par
ii.-The number of the isolating constants is $A+1.$%
\par
iii.-The number of the non isolating constant si $N-A.$%
\par
iv.-The number of configuration coordinates is $N+1.$%
\par
v.- The dimension of ${\cal M}$ is $2N-A+1$}. The $J_{j}^{W}$ can only be
considered as {\it local coordinates} on the manifold ${\cal M}%
(r_{0},...,r_{A}).$

When $A=N$ we say that the system is integrable, when $A<M$ we say that the
system is not integrable. Let us consider both cases.

\subsubsection{Integrable systems.}

In this case all the $P$ are isolating constants of the motion ($H)$ and
there are no $J.$ Then the situation is like the one described at the end of
section III: condition (\ref{3.3}) foliates phase space with submanifolds $%
{\cal M}(r_{0},...,r_{N}),$ labelled by the constants $(r_{0},...,r_{N}).$
This submanifold would be tori if the system is bounded. As we have already
said we will only consider the bounded case\footnote{%
The unbounded case requires a new singular structure in the states (see \cite
{CyLII}). This generalization will be studied elsewhere.}. In the tori the
motion of the configuration variables is given by eq. (\ref{5.10}). In the
generic case the $\varpi _{j}(\Pi )$ are not rationally dependant (or
non-commensurable). Then the trajectories of the configuration variable fill
each torus in a dense way. Therefore the motion is ergodic in each torus.
Moreover, we can independently see from eq. (\ref{otra}) that there is a
unique equilibrium state in each torus 
\begin{equation}
\rho _{*}^{W}(q,p)=\rho (r_{0},r_{1},...,r_{N})  \label{3.4'}
\end{equation}
which is constant in the submanifold. Therefore we have a microcanonical
equilibrium in each torus.

At this point we know that there is a unique stationary equilibrium state $%
\rho _{*}^{W}(q,p)$ $=\rho (r_{0},...,r_{N})$ but we do not even know if the
evolution converges to $\rho _{*}^{W}.$ So now we will prove, using the
quantum equations ( \ref{RO1}) or (\ref{5.2}), that when $t\rightarrow
\infty ,$ any $\rho ^{W}\rightarrow \rho _{*}^{W}$ in a weak way.

In the quantum case we have 
\begin{equation}
\lim_{t\rightarrow \infty }(\rho (t)|O)=(\rho _{*}|O)  \label{3.16}
\end{equation}
Now, quantum products can be changed into classical products according to
eq. (\ref{mean}) so: 
\begin{equation}
\lim_{t\rightarrow \infty }(\rho ^{W}(t)|O^{W})=(\rho _{*}^{W}|O^{W})
\label{3.17}
\end{equation}
and we have proved that: 
\begin{equation}
W\lim_{t\rightarrow \infty }\rho ^{W}(t)=\rho _{*}^{W}  \label{3.18}
\end{equation}
Therefore any $\rho ^{W}\rightarrow \rho _{*}^{W}$ weakly when $t\rightarrow
\infty .$

\subsubsection{Non integrable case.}

In this case not all the $P$ are isolating constants of the motion so there
are $H$ and $J.$ Then the trajectories must be dense in a domain ${\cal D}%
(r_{0},...,r_{A}){\cal \subset M}(r_{0},...,r_{A})$ of dimensions $2N-A+1.$
If not the dimensions of ${\cal D}(r_{0},...,r_{A})$ would be $<2N-A+1$%
\footnote{%
We may say that $A+1$ tori are not broken and that $A$ tori are broken.}$.$
In this case a new global constant must exist and there would be $A+2$
global constants. But this is impossible since $A+1$ is the total number of
these constants. It is quite clear that, as the equilibrium classical
density must be globally defined in ${\cal D}(r_{0},...,r_{A})$ the $%
J_{1}^{W}(q,p),...,J_{N-A}^{W}(q,p)$ cannot be explicit variables{\it \ }of $%
\rho _{*}^{W}(q,p)$\footnote{%
From its definition, in eq. (\ref{nonint}) or in eq. (\ref{5.7}), $\rho $ is
just an ordinary function of global variables. Therefore it cannot be
considered as a function defined using local coordinates.}$,$ so we must
just have: 
\begin{equation}
\rho _{*}^{W}(q,p)=\rho (H_{0}^{W}(q,p),\text{ }%
H_{1}^{W}(q,p),...,H_{A}^{W}(q,p))  \label{3.7}
\end{equation}
But on ${\cal D}(r_{0},...,r_{A})$ we have that: 
\begin{equation}
H_{0}^{W}(q,p)=r_{0}=const.,\text{ }%
H_{1}^{W}(q,p)=r_{1}=const.,...,H_{A}^{W}(q,p)=r_{A}=const.  \label{3.8}
\end{equation}
so on ${\cal D}(r_{0},...,r_{A})$ it is:

\begin{equation}
\rho _{*}^{W}(q,p)=\rho (r_{0},\text{ }r_{1},...,r_{A}).  \label{3.9}
\end{equation}
and we find the unique equilibrium for each ${\cal D}(r_{0},...,r_{A})$,
namely for each set of constants $(r_{0},...,r_{A}).$ Now we can follow the
reasoning of the previous section. Phase space is now foliated by
submanifolds ${\cal M}(r_{0},...,r_{A})$ of dimension $2N-A+1.$ The changes
are that now not all the coordinates of these submanifolds are configuration
variables\footnote{%
Regarding the configuration variables it is clear that, in the
non-integrable case, none of then is a global constant of the motion, that
further reduces the dimension of ${\cal M}$ (or ${\cal D).}$ In fact:
\par
i.- The preserve tori satisfy an irrationality condition (\cite{Tabor}, eq.
(3.4.12)) so the corresponding ratios of the frequencies are irrational and
the trajectories are dense in these tori.
\par
ii.- In the broken tori the trajectories are chaotic.
\par
Nevertheless, if in a particular case there is a configuration variable $X$
that turns out to be a global constant of the motion it can be considered
among the $"H"$. Then we will work essentially in the manifold $X=const.$
and nothing will change.}, there are also momentum variables, therefore we
cannot use the reasoning about the $\varpi _{j}(\Pi )$ not rationally
related in this case. Nevertheless in each ${\cal D}(r_{0},...,r_{A}){\cal %
\subset M}(r_{0},...,r_{A})$ there is a unique equilibrium state (\ref{3.9})
so using theorem 4.3 of ref. \cite{Mackey} we conclude that the motion is
ergodic in ${\cal D}(r_{0},...,r_{A}).$ Now we can repeat the reasoning of
eqs. (\ref{3.16}) to (\ref{3.18}) to show that the equilibrium (\ref{3.9})
is weakly reached. Since this equilibrium is a constant we again find a
microcanonical ensemble.

But, at this point we may ask ourselves why function $\rho
(H_{0}^{W}(q,p),H_{1}^{W}(q,p),...,H_{A}^{W}(q,p)J_{1}^{W}(q,p),...,J_{N-A}^{W}(q,p)) 
$ looses its $J$ variables. To explain this fact we may say that really
space ${\cal O}$ must contain physical measurable observables. But only the
CSCO can be measured in an independent way since their observables commute.
Moreover as the classical momenta $J_{1}^{W}(q,p),...,J_{N-A}^{W}(q,p)$ have
an ergodic motion so it is reasonable to consider that the quantum analogues 
$J_{1},...,J_{N-A},$ cannot really be measured, even at the quantum level%
\footnote{%
We can only measure dynamical variables when they can be considered as
constants in time, at least in the period of measurement. If a constant is
not global it is only constant in time in a local coordinate system, i. e.
it is not really physically constant.}. Then the set \{$H_{0},...,H_{A}\}$
is the relevant measurable CSCO, and it is only possible to measure the $%
H,H_{1},...,H_{A}$. Then the isolating constant \{$H_{0},...,H_{A}\}$
remains as the only characters in the quantum or classical play.

Phrased in another words the $J$ variables can only be diagonalized locally,
while the diagonalization procedure of section II.A.2 was thought as a
global one. So it is better to consider that the unitary operators $U$ of
eq. (\ref{2.11'}) only diagonalize the indices of the $H$ (that we will call 
$r$) and do not diagonalize the indices of the $J$ (that we will call $m).$
Then we obtain a basis where the coordinates of the stases read ( see (\ref
{nonint})) 
\begin{equation}
\rho (\omega )_{rmr^{\prime }m^{\prime }}=\rho _{rmm^{\prime }}(\omega
)\delta _{rr^{\prime }}  \label{3.10}
\end{equation}
Then eqs. (\ref{RO1}) and (\ref{RO2}) become 
\begin{equation}
\rho _{*}=W\lim_{t\rightarrow \infty }\rho (t)=\sum_{rmm^{\prime }}\int
d\omega \rho _{rmm^{\prime }}(\omega )(\omega ,rmrm^{\prime }|  \label{3.11}
\end{equation}
and 
\begin{equation}
P_{i}=\sum_{rmm^{\prime }}\int d\omega P_{rmm^{\prime }}^{i}|\omega
rm\rangle \langle \omega rm^{\prime }|,\qquad i=1,2,...,A  \label{3.12}
\end{equation}
and so on for the rest of the equations.

In order to follow this analysis we must ''trace away'' the $m$ since the
corresponding observables cannot we measured, i. e. they cannot be
considered classical in a global way. Essentially we must consider the $J$
operators and the $m$ indices as inexistent in space ${\cal O}$. As ${\cal O}
$ is the space of all measurable observables this fact must be considered as
the following change in the observables of eq. (\ref{2.5}) 
\begin{equation}
O(\omega )_{rmr^{\prime }m^{\prime }}\rightarrow O(\omega )_{rr^{\prime
}}\delta _{mm^{\prime }}  \label{3.13}
\end{equation}
(we consider only the diagonal term since we are only concerned in the
classical part and we neglect the $\omega _{0}$ term since we are
systemically forgetting the ground state). In this way the $m$ index has a
''spherical symmetry'' and they measure nothing. Then the relevant part of
eq. (\ref{2.8}) reads 
\[
\langle O\rangle _{\rho }=\sum_{rmr^{\prime }m^{\prime }}\int d\omega 
\overline{\rho (\omega )_{rmr^{\prime }m^{\prime }}}O(\omega )_{rmr^{\prime
}m^{\prime }}=\sum_{rmr^{\prime }m^{\prime }}\int d\omega \overline{\rho
(\omega )_{rmr^{\prime }m^{\prime }}}O(\omega )_{rr^{\prime }}\delta
_{mm^{\prime }}= 
\]
\begin{equation}
\sum_{rr^{\prime }}\int d\omega \left( \sum_{m}\overline{\rho (\omega
)_{rmr^{\prime }m}}\right) O(\omega )_{rr^{\prime }}  \label{3.14}
\end{equation}
Calling $\sum_{m}\overline{\rho (\omega )_{rmr^{\prime }m}}=\overline{\rho
(\omega )_{rr^{\prime }}\text{ }}$ we obtain the ''traced'' equation: 
\begin{equation}
\langle O\rangle _{\rho }=\sum_{rr^{\prime }}\int d\omega \overline{\rho
(\omega )_{rr^{\prime }}\text{ }}O(\omega )_{rr^{\prime }}  \label{3.15}
\end{equation}
and the $m$ indices have disappeared. From now on we can work with only the $%
r$ indices $(r_{0},...,r_{A})$ and in this way the $\rho $ of eq. (\ref{3.4'}%
) becomes the $\rho $ of eq. (\ref{3.9}) solving the problem.

Two comments are in order:

i.- From what we have said eq. (\ref{RO1}) can be considered as the quantum
version of the microcanonical equilibrium. Therefore the conditions to
obtain decoherence are equal to those to obtain microcanonical equilibrium
states. But this state will be different according to the number of $J.$

ii.- $H$ has a continuous spectrum, so this is also the case of ${\Bbb L.}$
(see (\ref{ara2}). Thus, albeit the $O(\hbar /s)$ of eq. (\ref{ara1}), we
can conclude that the classical $L$ has a continuous spectrum. Moreover,
normally quantum corrections make discrete the classical continuous spectra,
and not viceversa. Even more, the spectrum of classical Liouville operators
are usually continuous. Then $L$, most likely, has a continuous spectrum.
This is one of the characteristic properties of mixing (and therefore
ergodic) systems, the spectrum of their evolution operators are continuous$.$
This is another way to see that the flow is mixing in ${\cal D}%
(r_{0},...,r_{A})$ \cite{RS}, so the existence of a weak limit is natural.

\subsection{The canonical and grand canonical ensemble.}

We have proved that our system reach a microcanonical equilibrium. Then, the
frequent presence of canonical equilibrium can be explained by at least
three reasonings:

{\bf i.}- {\bf The thermodynamic limit. }In this limit microcanonical and
canonical densities coincide (see e. g. \cite{Hill}).

{\bf ii.- Canonical subsystem of a microcanonical system. }A small subsystem
of a big microcanonical system (that can be considered as a thermic bath) is
canonical \cite{Hill}. So, if the only isolating constant of the motion is
the energy $H,$ we have reach to the notion of canonical equilibrium: 
\begin{equation}
\rho _{*}^W\sim e^{-\beta H^W}  \label{A.1}
\end{equation}
where $\beta =T^{-1}.$ In the more general case of section III we would
arrive to the conclusion that (cf. \cite{Ajiezek}) 
\begin{equation}
\rho _{*}^W\sim \exp (-\beta H_0^W-\gamma _1H_1^W-...-\gamma _AH_A^W)
\label{A.2}
\end{equation}
namely generalized grand-canonical ensembles.

{\bf iii.- Information theory. }If we only use {\it unbiased distributions}
as in section II.B{\it \ }the exponential will naturally appear \cite{Jaynes}%
. Moreover our CSCO are numerable set, even if some of its observables have
continuous spectrum, so we can use the formalism of the quoted paper \cite
{Jaynes}, based on the information theory, with only one modification: to
use, for the observables with continuous spectrum, the classical continuous
version of Shannon ${\cal H}[\rho _{*}]$ (eq. (\ref{J.4})). Then we will
again find eqs. (\ref{A.1}) or (\ref{A.2}).

In all these ways we can obtain the canonical ensemble, to define
temperature and to begin the development of classical thermodynamics. As at
the end of section II.B.2 we can say that the system will thermalize or not
according to the kind of its interaction.

We have a final verification: to relate the equations of II.B with the
classical equations above.

Let us begin with the ''trace of an operator'' eq. (\ref{J.7}) in the
simplified CSCO $\{H\}$. The classical object corresponding to $(\omega |$
is by eq. (\ref{5.6}) 
\begin{equation}
\rho _{\omega }^{W}(q.p)=\delta (H^{W}(q,p)-\omega )  \label{A.3}
\end{equation}
so the classical object corresponding to $(I|$ is 
\begin{equation}
I^{W}(q,p)=\int \delta (H^{W}(q,p)-\omega )d\omega =1  \label{A.4}
\end{equation}
Therefore the classical trace of an operator corresponding to $(I|O)$ is 
\begin{equation}
(I^{W}|O^{W})=\int O^{W}(q,p)dqdp  \label{A.5}
\end{equation}
namely the integral usually associated with the trace. Finally the classical
equation corresponding to (\ref{J.8}) reads 
\begin{equation}
w_{\beta }^{W}(A)=\frac{\int e^{-\beta H^{W}(q,p)}A^{W}(q,p)dqdp}{\int
e^{-\beta H^{W}(q,p)}dqdp}  \label{A.6}
\end{equation}
i.e. eq. (\ref{A.1}), if we consider $\rho _{*}^{W}$ as a functional of $A$
and for the simplified CSCO $\{H\}.$ In the general case the functional
version of (\ref{A.2}) would be 
\begin{equation}
w_{\beta }^{W}(A)=\frac{\int e^{-\beta H^{W}-\gamma _{1}H_{1}^{W}-...-\gamma
_{A}H_{A}^{W}}A^{W}(q,p)dqdp}{\int e^{-\beta H^{W}-\gamma
_{1}H_{1}^{W}-...-\gamma _{A}H_{A}^{W}(}dqdp}  \label{A.6'}
\end{equation}
i. e., the classical version of (\ref{J.11'}) restricted to the isolating
members of the CSCO.

\section{Localization.}

The fate of a classical statistical system is to remain unlocalized in the
phase space or to localize around a classical trajectory, in which case the
system becomes classical. This fate depends in the potential acting in the
system (which can be localizing potentials or not) and the initial
conditions. Let us consider a classical distribution $\rho (q,p)$ with
support of volume $\Delta V^{(N+1)}.$ It would localize if $\Delta
V^{(N+1)}\rightarrow 0$ when $t\rightarrow \infty .$ But this is impossible
since we are dealing with an hamiltonian system where, according to
Poincar\'{e} theorem, the volume of phase space remains constant.
Nevertheless we see localized classical objects. This is only possible of
the system corresponds to those studied in section IV. A. 3 where there is a
volume $\Delta V_{O}^{(A+1)},$ that corresponds to the projection of the
support of $\rho (q,p)$ over the subspace of the observed dynamical
variables $H$, that vanishes when $t\rightarrow \infty ,$ while the volume $%
\Delta V_{U}^{(N-A)}$ corresponding to the unobservable $J$ dynamical
variables diverges when $t\rightarrow \infty $ being 
\begin{equation}
\Delta V_{U}^{(N-A)}.\Delta V_{O}^{(A+1)}\sim \Delta V^{(N+1)}=const.
\label{Poincaré}
\end{equation}
Therefore the subsystem of the observable variables is not hamiltonian and
the Poincar\'{e} theorem is by-passed. We conclude that only non integrable
system become classical and that the ''non-observability'' invoked in
section IV.A.3 is essential to obtain a final classical mechanical limit%
\footnote{%
We will give examples of the localization phenomenon elsewhere.}

\section{Conclusion.}

In usual text books of thermodynamic or statistical mechanics the ergodicity
of the system or the micro-canonical state are postulated. With our method
we have proved these postulates for systems with evolution operator endowed
with continuous spectrum only\footnote{%
If they decohere, namely if they satisfy the condition stated in paper \cite
{Deco}.}. Is this a limitation of our result? Quite on the contrary, it
characterizes the systems where thermodynamic and statistical mechanics can
be used: ergodic or mixing systems, being the latter one endowed with
continuous evolution spectra \cite{RS}. Of course if we work on a box the
evolution operator may have a discrete spectrum and we cannot reach to this
conclusion. But we know that in this case sooner or later we must expand the
box to infinity. Then, either we use hand waving argument: e. g. that the
distance between the values of the spectrum is very small etc., or we use
the rigorous result that in the limit the spectrum is continuous, and from
this fact deduce we the physical properties of the system, as we have done
in this paper


\begin{references}
\bibitem{Ant}  I. Anatoniou, Z. Suchanecki, S. Tasaki, R. Laura. Physica A, 
{\bf 241, }737, 1996.

\bibitem{CyLII}  R. Laura, M. Castagnino, Phys. Rev. E, {\bf 57, }3948, 1998.

\bibitem{Fund}  R. Laura, M. Castagnino{\it ,} Phys. Rev. A, {\bf 57},
4140-4152,1998.

\bibitem{vH}  L. van Hove, Physica, {\bf 23,} 268, 1979.

\bibitem{Deco}  M. Castagnino, R. Laura, Phys. Rev. A., {\bf 62}, 022107,
2000.

\bibitem{Splach}  S. H. Strogatz, {\it Non linear dynamics and chaos,}
Addison-Wesley, Reading, 1994.

\bibitem{Ballentine}  L. E. Ballentine,{\it \ Quantum mechanics,} Prentice
Hall, Englewoods Cliffs, 1990.

\bibitem{Bogo}  N. N. Bogolubov, A. A. Logunov, I. J. Todorov, {\it %
Introduction to axiomatic quantum field theory,} Benjamin, London, 1975.

\bibitem{CyLI}  M. Castagnino, R. Laura, Phys. Rev. A, {\bf 58,} 108-119,
1997.

\bibitem{CyLIII}  Castagnino M., Gunzig E.{\it , }Int. Journ. Theo. Phys., 
{\bf 38, }47, 1999.

\bibitem{GyHUCS}  M. Gell-Mann, J. Hartle, USCSRTH-94-09, 1994.

\bibitem{Zurek}  W. H. Zurek, {\it Preferred sets of states, predictability,
classicality, and environment- induced decoherence. In ''Physical Origin of
Time Asymmetry}'', Halliwell J. J. et al. eds., Cambridge University Press,
Cambridge, 1994.

\bibitem{Jaynes}  E. T. Jaynes, Phys. Rev., {\bf 106,} 620, 1957.

\bibitem{Katz}  A. Katz, {\it Principles of statistical mechanics, }Freeman,
San Francisco, 1967.

\bibitem{L&L}  L.D. Landau,E. M. Lifshitz, {\it Statistical physics, }%
Pergamon Press, Oxford, 1958

\bibitem{Haag}  R. Haag, {\it Local quantum physics, }Springer-Verlag,
Berlin, 1993.

\bibitem{Wigner}  M. Hillery, R. F. O'Connell, M: O.Scully, E. P. Wigner,
Phys. Rep., {\bf 106,} 123, 1984.

\bibitem{Mackey}  M. C. Mackey, {\it Time's arrow: the origin of
thermodynamics behavior, }Spinger-Verlag, Berlin, 1992.

\bibitem{D. Bhom}  D. J. Bhom, B. J. Hiley, Foundations of Physics, {\bf 5,}
93, 1975.

\bibitem{Arnold}  V. I. Arnold, {\it Mathematical methods of classical
physics, }Springer-Verlag, Berlin, 1973.

\bibitem{Tabor}  M. Tabor, {\it Chaos and integrability in non-linear
dynamics. }John Wiley \& Sons, New York, 1989.

\bibitem{Balescu}  R. Balescu, {\it equilibrium and non-equilibrium
statistical mechanics, }John Wiley \& Sons, New York, 1975

\bibitem{RS}  M. Reed, B. Simons, {\it Methods of modern
mathematical-physics III,} Academic Press, New York, 1979{\it \ }

\bibitem{Hill}  T. L. Hill, {\it An introduction to statistical
thermodynamics,} Dover Publications, New York, 1986.

\bibitem{Ajiezek}  A. I. Akhiezer, S. V. Peletminskii, {\it Methods of
statistical physics, }Pergamon Press, Oxford, 1981.
\end{references}
\end{document}